\documentclass[12pt]{iopart}

\usepackage{epsfig}
\usepackage{color}
\usepackage{graphics}
\usepackage{amssymb}

\begin{document}

\title[Mixing and segregation of ring polymers]{Mixing and segregation of ring
polymers: spatial confinement and molecular crowding effects}

\author{Jaeoh Shin $^\dagger$, Andrey G. Cherstvy$^\dagger$, and Ralf
Metzler$^{\dagger,\sharp,1}$}
\address{$^\dagger$Institute for Physics \& Astronomy, University of Potsdam,
14476 Potsdam-Golm, Germany\\
$^\sharp$Department of Physics, Tampere University of Technology, 33101
Tampere, Finland}
\ead{$^1$rmetzler@uni-potsdam.de}

\begin{abstract}
During the life cycle of bacterial cells the non-mixing of the two ring-shaped
daughter genomes is an important prerequisite for the cell division process.
Mimicking
the environments inside highly crowded biological cells, we study the dynamics and
statistical behaviour of two flexible ring polymers in the presence of cylindrical
confinement and crowding molecules. From extensive computer simulations we
determine the degree of ring-ring overlap and the number of inter-monomer contacts
for varying volume fractions $\phi$ of crowders. We also examine the entropic
de-mixing of polymer rings in the presence of mobile crowders and determine the
characteristic times of the internal polymer dynamics. Effects of the ring length
on ring-ring overlap are also analysed. In particular, on systematic variation of
the fraction of crowding molecules a $(1-\phi)$-scaling is found for the
ring-ring overlap length along the cylinder axis, and a non-monotonic dependence
of the 3D ring-ring contact number is predicted. Our results help to rationalise
the implications of macromolecular crowding for circular DNA molecules in confined
spaces inside bacteria as well as in localised cellular compartments inside
eukaryotic cells.
\end{abstract}


\section{Introduction} 

The physical effects of spatial confinement on the properties of ring polymers
is important to the physical understanding of the entropy-driven segregation of
the two bacterial daughter chromosomes upon cell division \cite{wright10entropy}
and the structure of eukaryotic metaphase chromosomes \cite{hancock-eu-crowding,
mirny1,mirny2}.
For rod-like bacteria cells such as \emph{E.~coli}, \emph{Bacillus subtilis},
or \emph{streptobacillus\/} a directed motion and segregation of duplicated
chromosomes along the cell axis is detected after DNA replication, see, for
instance, Ref.~\cite{kleckner-ecoli}. In eukaryotes, upon decondensation of
the chromosomes in a strongly limited space inside the nucleus, the existence of
chromosomal territories \cite{cremercremer,stasiak09} indicates an ultra-slow
polymer mixing dynamics \cite{rosaeveraers08,rosaeveraers11,yuval,kroy}. Knotting
of DNA molecules in tight spaces inside viral capsids another example of external
polymer confinement in biology \cite{phage09,virus13}. \emph{In vitro}, the
elongation and compaction of long DNA molecules confined in nano-channels upon
increasing fraction of the crowding agent was indeed detected \cite{maarel}. 

Internal polymer confinement \emph{in vivo\/} is due to macromolecular crowding,
which enforces DNA condensation in bacterial cells \cite{zimm96,elk10} where the
volume fraction occupied by crowding macromolecules such as RNA, ribosomes, or
other biomacromolecules reaches $\phi=30\dots35\%$ \cite{minton93}. The
abundance of crowding agents effects a viscoelastic environment \cite{igor,jeon1}
that severely alters the diffusional dynamics of endogeneous cytoplasmic granules
and of submicron tracers inside living
cells \cite{pt,golding,weber,weiss,jeon,tabei}. Concurrently the internal dynamics
of polymers and the macromolecular association kinetics inside biological cells
are dramatically changed \cite{zhou04,kapral12,marenduzzo13}.
The effect of various polymeric crowders on the opening-closing dynamics of DNA
hairpins has recently been experimentally probed in Ref.~\cite{weiss13crowd}.
Crowding can also
facilitate phase separation and compartmentalisation of the bacterial cytoplasm.
In theoretical models, inert spherical obstacles are often used to mimic highly
crowded interiors of bacterial \cite{minton93} and eukaryotic \cite{lang11}
cells. Crowding particles cause effective interactions between the polymer
segments of the same chain and between the two chains in confinement, as
studied in the present paper.

\begin{figure}
\begin{center}
\includegraphics[width=16cm]{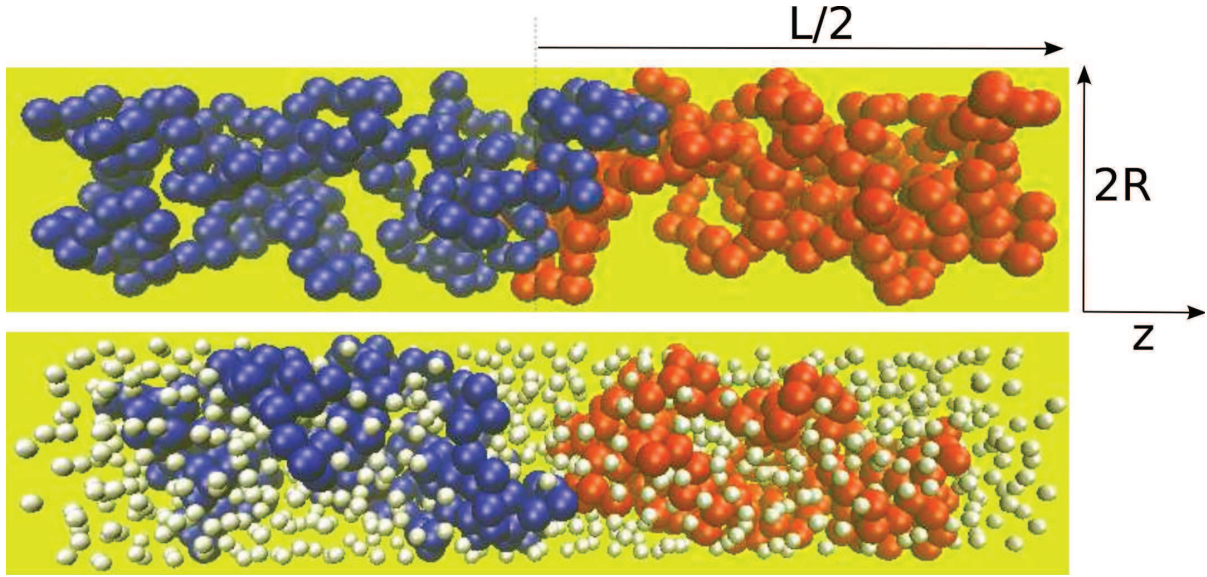}
\caption{Typical conformations of two polymer rings (rendered in red and blue,
respectively) in a cylindrical confinement. Parameters: the rings are $n=200$
monomers long, the cylinder radius is $R=4.5\sigma$ in terms of the monomer size
$\sigma$, with the fraction of crowders $\phi=0$ (top) and $\phi=0.182$ (bottom).
Crowding particles, which are equal in size to the chain monomers, are shown in
this image in light-yellow colour with 1/2 of their actual radius, to improve the
visibility of the image.}
\label{fig-scheme-chains}
\end{center}
\end{figure}

From the theoretical perspective, overlapped segments of long polymer chains
experience entropic repulsion scaling with the number of overlapping polymer
blobs \cite{jun10}. In a dense polymer melt the entanglements of the chains
also slow down the polymer dynamics \cite{mulder-repulsion,metzler02links}. In
the presence of obstacles, the extension and dynamics of ring polymers on
the lattice was analysed in Refs.~\cite{rubin-rings-obstacles}.   

A number of simulations studies of polymers under external confinement in various
geometries appeared in the literature in recent years \cite{ital-plane,
ital-sphere,ital-review,koreans12,likos12,reich13rings}. In particular, the
size scaling of ring polymers in dense melts was analysed by computer simulations
in Ref.~\cite{reich13rings}. As the concentration of rings $c$ grows and the
effective volume available for their expansion decreases, the scaling exponent for
the radius of gyration $\left< R_g^2\right>\simeq n^{2\nu}$ decreases from $\nu=3/5$
to $\nu\approx0.3$, mirroring impeded polymer extension. Neighbouring rings in dense
melts thus induce a spherical caging effect, and their dimension was shown to scale
as $\left< R_g^2 \right>\sim c^{-0.59}$ in terms of the ring concentration
\cite{reich13rings}. The segregation of semi-flexible macromolecules in
nano-channels was shown theoretically in Ref.~\cite{slovak13}.
Ring polymers in confinement were successfully used to model the bacterial
chromosome \cite{koreans12} and to rationalise the implications of supercoiling
for the contact maps of eukaryotic interphase chromosomes \cite{stasiak14chrom-NAR}.

More specific with respect to our present study, the entanglement propensity of
ring and linear polymers under external cylindrical confinement with respect to
the phenomenon of DNA homologous recombination were analysed recently in
Ref.~\cite{stasiak13}. It was shown that linear chains penetrate into one another
significantly easier than ring polymers. Finally, the threading of ring polymers
inside a polymer gel was recently studied by simulations \cite{italy-gel-rings}.

The statistical properties of linear and ring polymers in the presence of crowding
effects were considered in a number of theoretical and simulations studies. In
particular, the behaviour of single knotted polymer rings on a regular lattice of
obstacles was simulated in Refs.~\cite{orlandini10pre,note-ring-RG-scaling}. For
random-loop and self-avoiding polymers in the presence of crowding the computer
modelling in Refs.~\cite{schzleifer11,schzleifer10} was shown to give rise to a
non-monotonic dependence of $\left< R_g^2\right>$ on the volume fraction occupied
by the crowders, $\phi$, featuring a slight minimum in the chain dimensions at
$\phi\approx0.2$. Related to this, the rates of chemical diffusion-limited
reactions in molecularly crowded media in confined environments was shown to
reveal a maximum at $\phi\approx0.2$ \cite{kondev09}. 

In a biophysical context, the effect of crowding on gene regulation was studied
with respect to facilitated diffusion and target search on DNA by DNA-binding
proteins in Refs.~\cite{marenduzzo13,levi-crowd}. The translocation of polymers
between the two reservoirs with crowders of equal \cite{kim07transl} and non-equal
\cite{liu-crowd} sizes has also been rationalised by simulations. The implications
of crowding environments inside nano-channels have also been recently examined by
simulations \cite{tere13}. Finally, the effects of crowders on the looping
probabilities of polymer chains in the presence of external confinement and
molecular crowding is also important \cite{shin-looping}.

Below, we take a step further and analyse the joint effect of external confinement
and internal crowding for two unknotted ring polymers in a model rod-shaped
bacteria. More
concretely, we perform molecular dynamics simulations for two polymer rings
confined in a cylindric volume in the presence of mobile crowders, which are
subject to the same thermal bath, as illustrated in Fig.~\ref{fig-scheme-chains}.
We analyse how the entropic repulsion of these thermally agitated ring polymers
becomes altered under these crowding conditions. 

The paper is organised as follows. In the next Section, we present the details
of the simulations model. In Section \ref{sec-results} we rationalise the effects
of external cylindrical confinement and internal confinement by the crowding
obstacles. The main results for the static properties of the mutual overlap of
the polymer rings and their dynamic characteristics are presented. In Section
\ref{sec-conclusions} we discuss the basic results and their implications to
the biological system and with respect to polymer physics.

\section{Model and implementation of the simulations}

\subsection{Polymer chains}

The standard finitely extensible non-elastic (FENE) potential is used to model
the interactions between the monomers in our polymer chains in the bead-spring
coarse-grained model of the DNA molecule, namely,
\begin{equation}
U_{\mathrm{FENE}}(r)=-\frac{k}{2}r_ {\mathrm{max}}^2\ln\left(1-\frac{r^2}{r_{
\mathrm{max}}^2}\right).
\end{equation}
Here $k$ is the spring constant acting between nearest-neighbour beads and $r_{
\mathrm{max}}$ is the maximum allowed separation between neighbouring monomers. The
total number of monomers in the ring polymers varies in the simulations in the
range $n=60\dots350$. Excluded-volume interactions between the polymer segments
are introduced by the truncated Lennard-Jones repulsion (Weeks-Chandler-Andersen
potential), that is,
\begin{equation}
U_{\mathrm{LJ}}(r)=\left\{\begin{array}{ll}4\epsilon[(\sigma/r)^{12}-(\sigma/r)^{6}]
+\epsilon, & r<2^{1/6}\sigma\\[0.2cm]
0, & \mbox{ otherwise}\end{array}\right.
\end{equation} 
We here introduced the monomer-monomer distance $r$, $\sigma$ is the monomer
diameter, and $\epsilon$ is the strength of the potential. We set $k=30$, $r_
{\mathrm{max}}=1.5\sigma$, and $\epsilon=1$, where all energies are measured in
units of the thermal energy $k_BT$, and distances are measured in units of
$\sigma$. Analogous repulsive 6-12 Lennard-Jones
potentials parameterise the chain-obstacle, chain-wall, and crowder-wall contact
interactions. Along the axial $z$ coordinate in our cylindrical geometry a harmonic
potential is applied once a monomer attempts to move outside of the cylinder at
$z<0$ or $z>L$. 

The dynamics of the position $\mathbf{r}_i(t)$ of the $i$th monomer in a polymer
chain is described by the Langevin equation
\begin{eqnarray}
\nonumber
m\frac{d^2\mathbf{r}_{i}(t)}{dt^2}&=& -\sum_{j=1,j\neq i}^{n}\nabla[U_{\mathrm{LJ}}
(\mathbf{r}_i-\mathbf{r}_j)+U_{\mathrm{FENE}}(\mathbf{r}_i\mathbf{-r}_j)]\\
&&-\sum_{j=1}^{N_{cr}}\nabla U_ {\mathrm{LJ}}(\mathbf{r}_i-\mathbf{r}_{cr,j})-
\nabla U_ {\mathrm{LJ}}(\mathbf{r}_i-\mathbf{R}_{\mathrm{cyl}})-\xi\mathbf{v}_i(t
)+\mathbf{F}(t).
\end{eqnarray}
Here $m$ is the monomer mass, $\xi$ is the friction coefficient, $v_i$ is the
monomer velocity, and $F(t)$ represents Gaussian delta-correlated noise with
$\delta$-correlations $\left<F(t)F(t')\right>=6\xi\delta(t-t')$. Similar Langevin
equations are used for the dynamics of the positions $\mathbf{r}_{cr,j}$ of the
crowding molecules in the presence of the confining cylinder at $\mathbf{R}_{
\mathrm{cyl}}$. Similarly to the procedure described in Ref.~\cite{scr13virus},
we implement the velocity Verlet algorithm with the characteristic integration
time-step of $\Delta t=0.01$. 

The monomer size is set to $\sigma=4$ nm determining the chain thickness that
stays constant for different ring lengths simulated below. This thickness
represents the effective physical DNA diameter including hydration water shells
and electrostatic effects \cite{voloDNAdiameter2}. Our approach thus
differs from that taken in Ref.~\cite{stasiak13}, where the mixing of ring
polymers of different lengths without a crowding agent was studied and the
polymers were assumed to become thinner as they get longer. This assumption was
used in Ref.~\cite{stasiak13} to keep a constant volume fraction $\phi_p$ of the
polymer chains $V_p$ in the simulation box of volume $V$, and it is estimated that
\begin{equation}
\phi_p=\frac{\mbox{DNA volume}}{\mbox{\emph{E. Coli} volume}}=\frac{V_p}{V}\sim1
\dots5,
\end{equation}
depending on the DNA thickness (the bare DNA diameter plus the electrostatic
repulsive salt-dependent shell around it). We present the ring-ring
contact number and overlap distance of rings for the fixed $\phi_p$ in
Fig. \ref{fig-contacts-final}.

The equilibration time of the polymer rings in our simulations depends on their
length, the cell cylinder radius $R$, and the volume fraction $\phi$ of crowding
particles in the simulation box. For typical parameters of the ring length
\begin{equation}
l=n\sigma\sim200\sigma
\end{equation}
and $R=4.5\sigma$ used in simulations below, the chains equilibrate after $\sim4
\times10^6$ simulation steps in the absence of crowders. The ring equilibration
time grows with the chain length, and longer measurement times are required
in order to sample the conformations of the polymer chains. The equilibration
time also grows with the volume fraction $\phi$ of crowders due to the
slow-down of the polymer dynamics, see below.

\subsection{Crowding and confinement}

We distinguish two types of volume confinement for the polymer chains: external
confinement by the cylindrical cell walls and internal confinement by mobile
crowding obstacles. The model cell in our simulations is represented by an
impenetrable cylinder of length $L=35\sigma$ and radius $R=3.5\sigma\dots5.5
\sigma$.

By the internal confinement we mimic the highly poly-disperse soup of various
proteins, RNA, cytoskeletal elements, and organelles in the cell cytoplasm. The
crowders in the bacterial cytoplasm have an average molecular weight of $\mathrm{
MW}\approx40\dots67$ kDa and diameter of 4$\dots$8 nm \cite{schzleifer11}. We
neglect here the poly-dispersity in crowder sizes observed in real cells
\cite{elcock-ecoli-cytoplasm} and for simplicity assign to the crowders the same
size $\sigma_{cr}=\sigma$ as for the chain monomers. The crowders are simulated
as spherical particles of unit mass (similarly to the polymer bead), with
systematically varying volume occupancy $\phi$. Each polymer monomer therefore
corresponds to $\approx12$ base pairs of the double-stranded DNA and $\mathrm{MW}
\approx12\times0.66$ kDa. Every step in our simulations then corresponds to a real
time of $\tau_0=\sigma\sqrt{12\times660~\mathrm{Da}/(k_BT)}\approx0.23$ ns.
Simulating the crowders as particles with the more realistic value of MW of 67 kDa
will slow down the crowder and polymer dynamics,
renormalising the elementary simulation time unit to $\tau_0\approx0.66$ ns.

Varying the volume fraction of crowders in the simulations in the range $0<\phi
\lesssim0.3$ we mimic the response of a cell to the changes in external osmolarity,
exerting a pressure on the outer cell membrane causing dehydration (osmotic
upshift) \cite{upshift}. This volume fraction is computed per free solution
volume, i.e.,
\begin{equation}  
\phi=\frac{V_{cr}}{V-V_{p}}=\frac{N_{cr}v}{\pi R^2 L-nv},
\end{equation}
where $v=4\pi(\sigma/2)^3/3$ is the volume of one chain monomer or of one crowding
particle and $N_{cr}$ is the number of crowding particles in the box. For
the chain length $n=200$
and cell length $L=35\sigma$ the volume fraction of the two polymer rings is
$\phi_p\approx0.155$, $0.094$, and $0.063$ (close to the DNA crowding in
\emph{E.coli\/} \cite{stasiak13}) for the respective cylinder radii
$R=3.5\sigma$, $4.5\sigma$, and $5.5\sigma$.

We consider only excluded volume interactions according to the above-mentioned
interaction potentials and neglect other interactions within the ring polymers,
including electrostatic interactions. The latter can be of importance for
tightly bent and circular DNAs, particularly at low-salt, weak-screening
conditions \cite{dna-elst,ac11rings,ac11rings2}. Our model also neglects effects
of hydrodynamic interactions (both for rings and crowders)
\cite{skolnick-hydro,wensink-hydro}, which can alter short-time polymer
dynamics \cite{china-rings-hydro}, but should not affect the static overlap
properties of the rings. Polymer relaxation under confinement with and without
hydrodynamic interactions was studied by computer simulations
\cite{korea09scaling}. For the relaxation time $\tau_R$ of a polymer ring
consisting of $n$ monomers in a long cylindrical pore of radius $R$ the relation
$\tau_R\sim n^2R^{0.9}$ was predicted \cite{koreans12,korea09scaling}.

\subsection{Ring contacts and decay of correlations}

For each simulation step $t$ we determine the number of contacts $N_{AB}(t)$
between the two ring polymers as follows. Each monomer is surrounded by a sphere
with contact radius $r_c=1.25\dots2\sigma$ that defines the overlap volume. If
the centre of mass of a monomer of another chain stays within this contact sphere
for the contact time $t_c$, the contact is recorded as established. The time $t_c$
is a measure of the internal dynamics of two intermingled rings. Within this time
scale, the change in distances between contacting monomers should be smaller than
$r_c$. This validates the choice of the temporal and spacial thresholds for
counting the number $N_{AB}$ of ring-ring contacts.
 
The distance $r_c$ represents the `radius of action' within which the monomers
are supposed to be involved in some physical interactions. For the DNA,
this can be electrostatic or protein-mediated inter-molecular contacts
\cite{prot-dna}. Clearly, the results of counting the number $N_{AB}$ of contacts
depends on the threshold distance $r_c$ (for comparison, the choice of $r_c=1.5
\sigma$ was used in Ref.~\cite{stasiak13}). We analyse the dependence of the
contact number on the contact distance $r_c$ in Fig.~\ref{fig-rc-all} below.

The average $\left<N_{AB}\right>$ is computed via averaging over various polymer
configurations after the system reached its equilibrium. The contact volume
$V_{AB}$ of the rings is estimated as $\left<N_{AB}\right>$ multiplied by $\sim
1/2$ of the volume of the contact sphere, $\left<V_{AB}\right>\approx\left<N_{AB}
\right>4\pi(r_c/2)^3/3$. In addition to the three-dimensional inter-chain contact
probability, scaling with $\left<N_{AB}\right>$, we compute the one-dimensional
mutual overlap length of two rings along the $z-$axis of the confining cylinder,
$\left<l_{AB}\right>$.

Note that we consider only torsionally relaxed rings, with no effects of
super-coiling. The latter would result in more branched and topologically
complex polymer structures, likely with more extensive contacts.

Following Ref.~\cite{stasiak13}, we define the auto-correlation function (ACF) of
ring-ring contacts via the contact number as follows
\begin{equation}
\label{eq-ACF}
\mathrm{ACF}(\Delta)=\frac{\left<N_{AB}(t+\Delta)N_{AB}(t)\right>-\left<N_{AB}(t+
\Delta)\right>\left<N_{AB}(t)\right>}{\left<N_{AB}(t)^{2}\right>-\left<N_{AB}(t)
\right>^2}.
\end{equation}
The averaging $\left<\dots\right>$ is performed over the times times $t$ along the
generated trace $N_{AB}(t)$ with the corresponding lag time $\Delta$. The ACF
characterises the decay of correlations in the
overlap number of rings. The equilibration time in all our simulations is at least
50 times longer than the correlation time of the corresponding ACF for the
inter-chain contact number for the chosen parameters.

An additional quantity characterising the ring-ring overlap is the relative
position of their centres of mass,
\begin{equation}
\Delta z_{CM}=z_{CM,A}-z_{CM,B}.
\end{equation}
From the corresponding probability density $p_{AB}(\Delta z_{CM})$ along the
cylinder axis we compute the
free energy of the overlap of the two rings in terms of
\begin{equation}
F(z_{CM})=-k_BT\log[p_{AB}(z_{CM})]
\end{equation}
in the Shannon sense.

\section{Results}
\label{sec-results}

\subsection{Dimensions and contacts of polymer rings}

\begin{figure}
\begin{center}
\includegraphics[width=8cm]{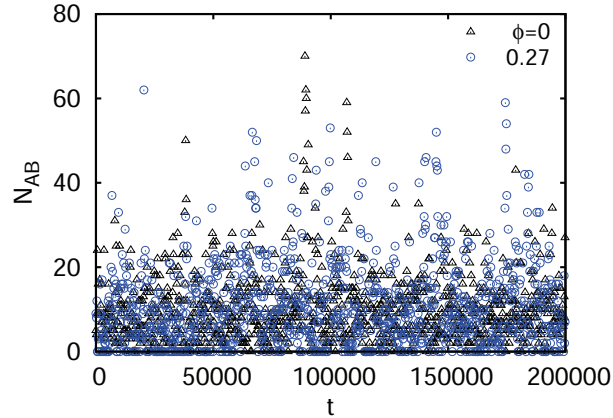}
\caption{Fluctuations of the contact number $N_{AB}(t)$ of polymer rings as
function of simulation time $t$, plotted for crowding fractions $\phi=0$ and
0.27. Parameters: the number of monomers in each ring is $n=200$, the radius
of the confining cylinder is $R=4.5\sigma$, the critical contact distance is
$r_c=1.5\sigma$, and the cylinder length is $L=35\sigma$. Here and below the
simulations time is presented in units of the time step $\delta t$.}
\label{fig-NAB-crazy}
\end{center}
\end{figure}

\begin{figure}
\begin{center}
\includegraphics[width=16cm]{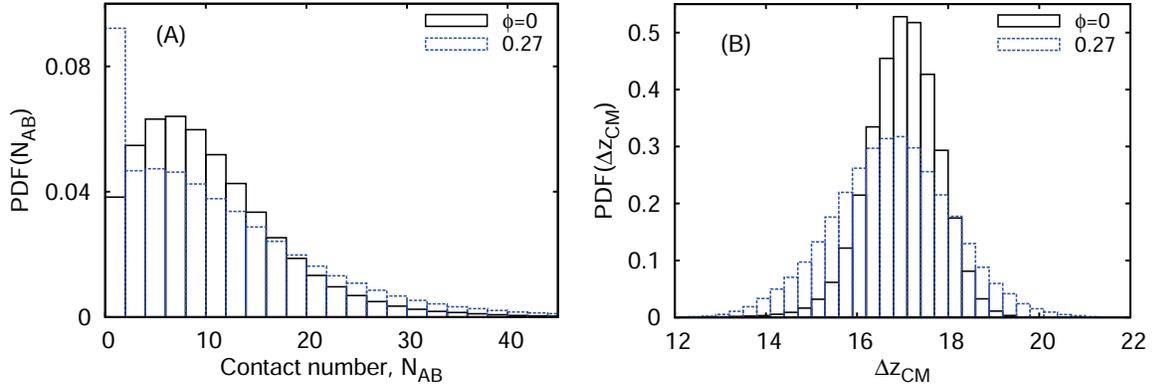}
\caption{Probability density function of the ring-ring contact number (A) and
centres of mass difference of the two rings (B), plotted for the parameters of
Fig.~\ref{fig-NAB-crazy}.}
\label{fig-n-pdf}
\end{center}
\end{figure}

We verified that the extension of an unconfined ring polymer scales with its
length as
\begin{equation}
\left<R_{g}^2(n)\right>\sim n^{6/5},
\end{equation}
consistent with the results reported in Ref.~\cite{orlandini10pre} (not shown).
Our polymer rings are very flexible, the effective persistence length $l_p$ being
of the
order of the monomer size (data not shown). Because of this extreme chain
flexibility, we cannot analyse the implications of confinement onto the chain
persistence (as compared to Ref.~\cite{koreans12} where ring polymers were shown
to stiffen substantially in tight confinement).
Under the cylindrical confinement, the ring size $\left<R_g^2\right>$ naturally
reaches a saturation for long chains. Once the chain dimensions overcome the
size of the cylindrical cell, the polymer starts to folds on itself and its
apparent scaling exponent $\nu$ decreases.

The initial ring configurations at $t=0$ generated in the simulations are
well-separated, positioned at the opposite sides of the confining cylinder.
They exhibit a fast
initial relaxation followed by a roughly exponential relaxation dynamics. At the
later stages, when the polymers experience external confinement by the cylinder
and the other ring, a non-exponential relaxation dynamics sets in. The spectrum
of chain fluctuations in frequency space in the presence of external confinement
and crowding becomes altered as well.

The general trend is that the instantaneous number of ring-ring contacts $N_{AB}
(t)$ fluctuates strongly in the course of the simulations, compare
Fig.~\ref{fig-NAB-crazy}. This trend is the same as in recent simulations for a
similar system presented in the Supplementary Material of Ref.~\cite{stasiak13}. We
observe that in the presence of crowders the ring-ring separation becomes
more pronounced, and the probability density function $\mathrm{PDF}(N_{AB})$ of
their contact numbers exhibits a peak at $\left<N_{AB}\right>=0$. The spread of
$N_{AB}$ is slightly more localised in the presence of crowders, but both at
crowded and non-crowded conditions the distributions $p(N_{AB})$ have long tails,
as evidenced in Fig.~\ref{fig-n-pdf} A. The relative centre-of-mass position
of the two rings, $\Delta z_{CM}$, shows a larger spread in the presence of
crowders, see Fig.~\ref{fig-n-pdf}B.

\subsection{Ring swapping}

\begin{figure}
\begin{center}
\includegraphics[width=8cm]{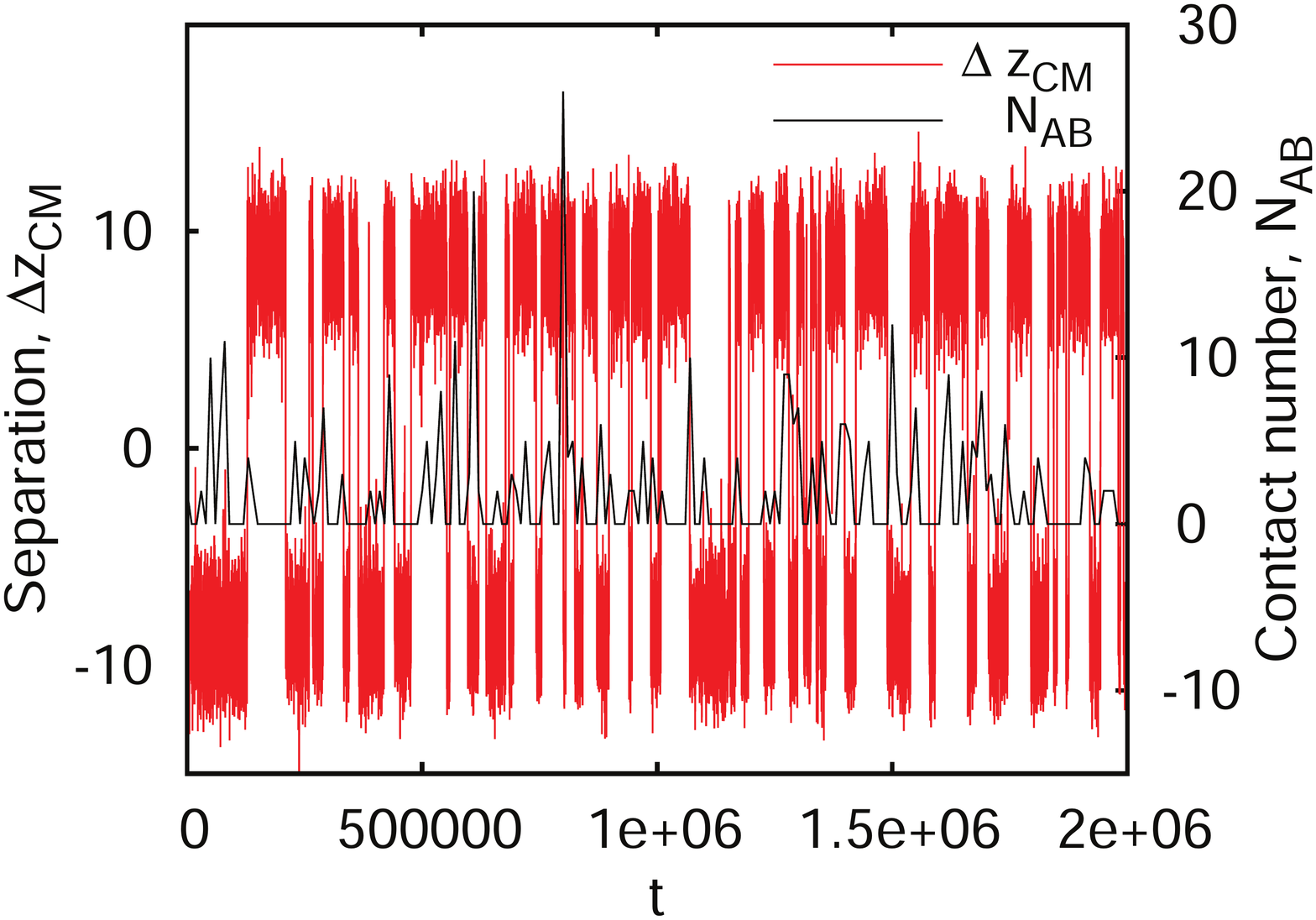}
\caption{Time traces for the contact number $N_{AB}(t)$ (black) and the centre of
mass distance of the rings $\Delta z_{CM}$ (red). Every 10,000th and 100th data
point is shown for the $N_{AB}(t)$ and $\Delta z_{CM}(t)$ trajectories,
respectively. Parameters: no crowders ($\phi=0$), chain lengths $n=60$, cylinder
radius $R=5.5\sigma$ and length $L=20\sigma$. A video illustrating the ring
swapping events is included in the Supplementary Material.}
\label{fig-NAB-XCM}
\end{center}
\end{figure}

\begin{figure}
\begin{center}
\includegraphics[width=8cm]{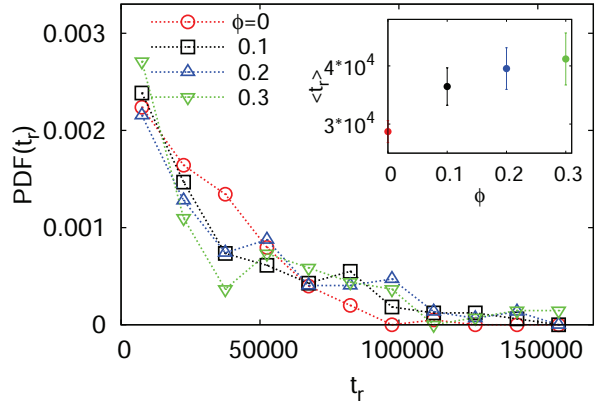}
\caption{The distribution of the residence times $t_r$ in the well-segregated
ring states. The histograms are obtained from the zigzag traces similar to
those presented in Fig.~\ref{fig-NAB-XCM}, but at varying crowding fractions.
The inset shows the mean residence time of rings $\left<t_r\right>$ in
well-separated states as a function of $\phi$.}
\label{fig-XCM-dist}
\end{center}
\end{figure}

For some choices of the volume and the aspect ratio of the confining cylinder as
well as for shorter polymer lengths, the directed distance $\Delta z_{CM}$ between
the centres of mass of the two ring polymers exhibits clear alternations between
two states while the rings are well separated near the ends of the confining
cylinder, see Fig. \ref{fig-NAB-XCM}. For such conditions, the diffusion times
of the rings along the cylinder are relatively short, so they can pass one another
and swap positions. At time instances when the rings are well
separated the number of ring-ring contacts is minimal, while at almost vanishing
centre of mass separation, $z_{CM}\sim0$, the overlap of the rings is maximal and
thus typically the $N_{AB}(t)$ traces are peaking at these instances, see the
time series of $\Delta z_{CM}(t)$ and $N_{AB}(t)$ shown in Fig.~\ref{fig-NAB-XCM}.
Note that the centre of mass distance $|\Delta z_{CM}|\leq L$, and the
case of $\Delta z_{CM}>0$ corresponds to the situation when ring A is located on
the left half and ring B on the right half of the confining cylinder. This type of
dynamics is reminiscent of the periodic tumbling of polymers in shear flows,
characterised by configurations with large extensions alternating with states of
strong chain contraction, see, e.g., the studies reported in
Refs.~\cite{winkler-shear,winkler-shear2}.

For these conditions of well separated rings, the distribution of residence times
$t_r$, that each ring is situated in one of the two well-separated states close to
the cylinder ends, is shown in Fig.~\ref{fig-XCM-dist} for different crowding
fractions $\phi$. The mean residence time $\langle t_r\rangle$ extracted from these
histogram is the characteristic time scale for the internal ring swapping dynamics.
As shown in the inset of Fig.~\ref{fig-XCM-dist}, $\langle t_r\rangle$ mildly
increases with increasing $\phi$. The maximum of the residence time histograms in
Fig.~\ref{fig-XCM-dist} shifts at
higher crowding fractions to larger values because of the associated slower
polymer dynamics.

\begin{figure}
\begin{center}
\includegraphics[width=7cm]{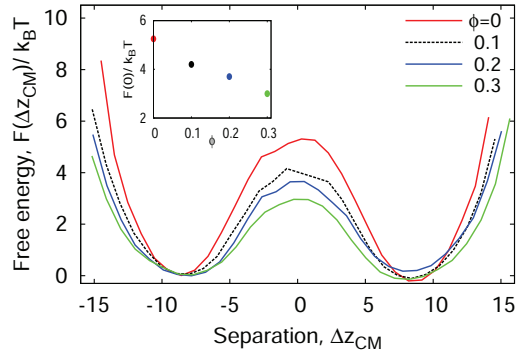}
\caption{Double-well free energy landscape for ring swapping. The inset shows
the magnitude of the free energy barrier for ring swapping computed at $\Delta
z_{CM}=0$. Parameters and notations for the curves are the same as in
Fig.~\ref{fig-XCM-dist}.}
\label{fig-FE-2wells}
\end{center}
\end{figure}

As illustrated in Fig.~\ref{fig-FE-2wells} the effective free energy for mixing
the two rings has a double-well shape. The height of the barrier separating the
two minimum states amounts to several $k_BT$ for the parameters used in the
simulations. As the residence times $t_r$ in these separated ring states
increases, the height of the free energy barrier between them decreases. This is
due to a slower polymer dynamics at high crowding fractions, as discussed below.
It also demonstrates that the free energy landscape is no true equilibrium
measure, as known from the theory of polymer translocation \cite{chuang}.

Longer rings squeezed into the same confining cylinder reveal a slower swapping
dynamics and the residence times in well-separated states grow until no swapping
at all can be observed during the simulation time. Likewise, the exchange of rings
in the simulation box is prohibited for smaller cylinder radii $R$ (not shown). 

\subsection{Correlations of ring-ring contacts}

\begin{figure}
\begin{center}
\includegraphics[width=7cm]{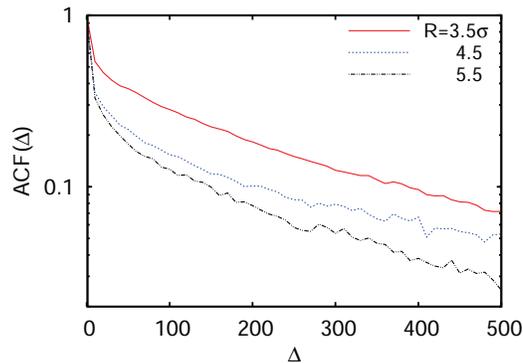}
\caption{ACF of the ring-ring contact number $N_{AB}$ defined in Eq.~(\ref{eq-ACF})
in absence of crowders ($\phi=0$) for different values $R$ of the cylinder radius.
Other parameters are the same as in Fig.~\ref{fig-NAB-crazy}.}
\label{fig-ACF}
\end{center}
\end{figure}

As demonstrated in Fig.~\ref{fig-NAB-crazy}, the number of ring-ring contacts
fluctuates strongly and irregularly. To find a typical time-scale for this
variation, we
compute the ACF (\ref{eq-ACF}) of the ring-ring contact number from the $N_{AB}(t)$
time traces. We start with the crowding-free case $\phi=0$. The resulting curves
in Fig.~\ref{fig-ACF} show a fast relaxation at short times and turn to a nearly
exponential decay at intermediate lag times $\Delta$. At long times, the ACF drops
to zero, indicating a complete loss of correlations. Some fluctuations of the
$\mathrm{ACF}(\Delta)$ at $\Delta\to\infty$ indicate insufficient statistics in
the calculation of the time average (\ref{eq-ACF}).
From Fig.~\ref{eq-ACF}, we observe that the
initial decay of the ACF is slower for smaller cylinder radii, as expected.
This is due to a larger space fraction in the  simulation box being filled by the
polymer monomers so that their motions get restricted to a larger extent, with many
chains' moves being prohibited. For longer rings confined in the same cylinder, the
ACF decays slower with the lag time $\Delta$, again due to a smaller space
available for the chains (not shown). Note that the intermediate-time decay
exhibits comparable slopes in the logarithmic plot of Fig.~\ref{eq-ACF}.

\subsection{Contacts and overlap of polymer rings: crowding effects}

\begin{figure}
\begin{center}
\includegraphics[width=14cm]{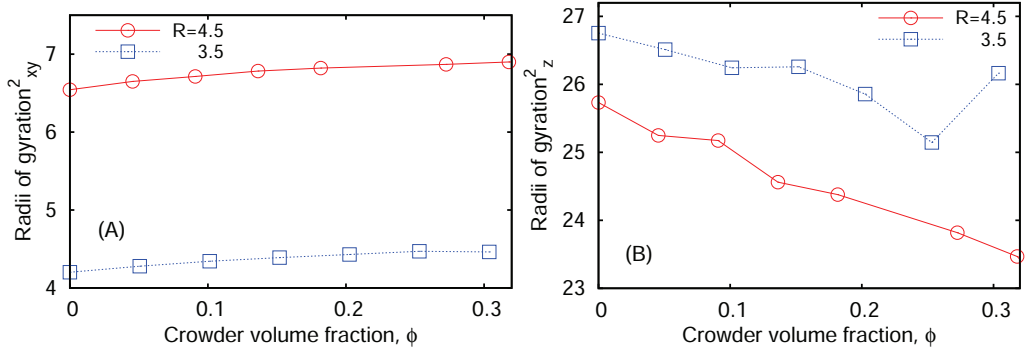}
\caption{Mean squared gyration radius $\langle R_g^2\rangle$ of a single polymer
ring across the confining cylinder (A) and along (B) the cylinder axis, computed
for varying crowding fractions $\phi$ and two different radii $R$ of the cylinder.
Other parameters are the same as in Fig.~\ref{fig-NAB-crazy}.}
\label{fig-RG2-single}
\end{center}
\end{figure}

Let us now study the effects of the internal confinement due to crowding in more
detail. We first consider a single ring polymer under the cylindrical confinement
in the presence of crowding agents. The results for the mean squared gyration
radius $\langle R_g^2\rangle$ are shown in Fig.~\ref{fig-RG2-single}. We observe
that the component of the radius of gyration measured along the cylinder axis is
a slowly decreasing function of $\phi$. Crowding particles thus act as a depletant,
that effects ring shrinkage. For a less severe external confinement (larger
cylinder radius $R$), we observe that the ring is more confined along the
cylinder axis, but simultaneously more extended in the cylinder cross-section
($x-y$ plane), as shown in Fig.~\ref{fig-RG2-single}. Here we do not elaborate
on the variation of the Flory scaling exponent $\nu$ of the gyration radius for a
single ring as function of the external confinement and crowding (for such results
see, e.g., the results reported in Ref.~\cite{note-ring-RG-scaling}). In the
following we concentrate on the overlap properties of two polymer rings in the
cylindrical simulations cell.

\begin{figure}
\begin{center}
\includegraphics[width=8cm]{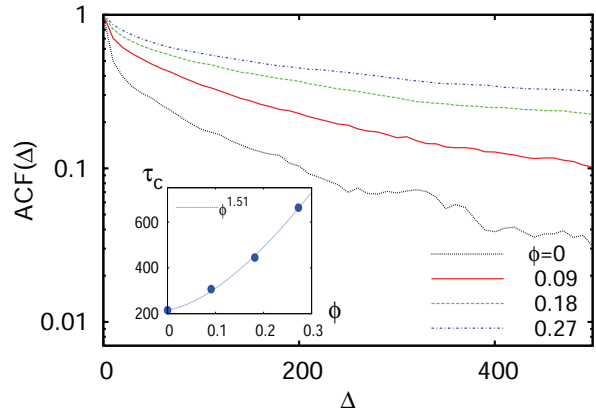}
\caption{ACF function of ring-ring contacts, Eq.~(\ref{eq-ACF}), for different
crowding fractions $\phi$ and the parameters of Fig.~\ref{fig-NAB-crazy}. The
inset shows the correlation time $\tau_C(\phi)$ determined from the decay
$\exp(-\Delta/\tau_C)$ of the $\mathrm{ACF}(\Delta)$ curves.}
\label{fig-acf-crowd}
\end{center}
\end{figure}

At higher fractions $\phi$ of crowders the correlation time of maintainig the
established contacts
between polymer rings increases due to the slower polymer dynamics, following a
larger effective viscosity in a denser soup of crowders, i.e., the Rouse polymer
dynamics becomes effectively slowed down by surrounding crowding particles.
Concurrently, the same effect is responsible for a slower decay of the contact
autocorrelations at higher values of $\phi$, as shown in Fig.~\ref{fig-acf-crowd}.
The associated correlation
time $\tau_C$ extracted from an exponential fit of the $\mathrm{ACF}(\Delta)$
curves exhibits the power-law dependence
\begin{equation}
\tau_C(\phi)\sim\phi^{3/2}
\end{equation}
on the fraction of crowders, see the inset of Fig.~\ref{fig-acf-crowd}. The $3/2$
exponent indicates that the changes due to crowding are indeed a volume effect.
Note that the single-ring relaxation time $\tau_R$ should not be confused with
$\tau_C$ for the ring-ring contacts.

\begin{figure}
\begin{center}
\includegraphics[width=14cm]{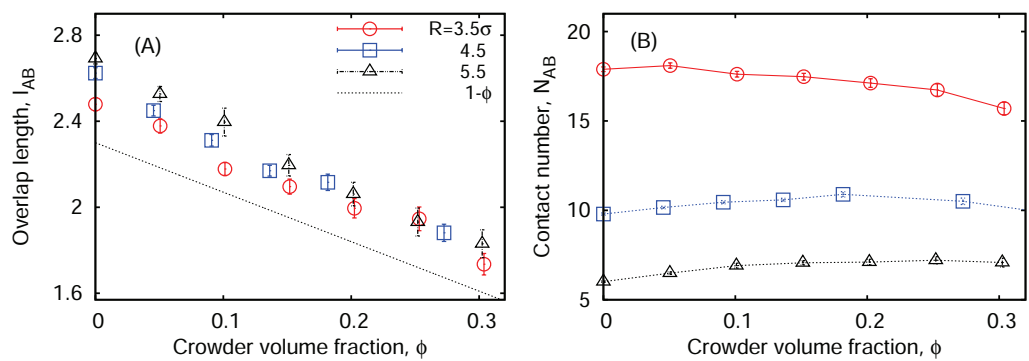}
\caption{Average ring-ring overlap length along the cylinder axis (A) and average
contact number (B), plotted for the parameters of Fig.~\ref{fig-NAB-crazy}.}
\label{fig-contacts-1D-crowding}
\end{center}
\end{figure}

The dependence on the presence of mobile crowders on the three-dimensional contact
characteristics and the effective one-dimensional overlap properties of polymer
rings of fixed length are analysed in Fig.~\ref{fig-contacts-1D-crowding}. We
observe that the average number of ring-ring contacts $\left<N_{AB}\right>$
increases significantly with the decrease of the cylinder radius $R$, i.e., when
the chains are forced into a stronger contact by the external confinement, see
Fig.~\ref{fig-contacts-1D-crowding}A. As
function of the internal confinement due to crowding, in some situations the
number of ring-ring contacts $N_{AB}$ exhibits a weakly non-monotonic dependence,
see, e.g., the blue symbols in Fig.~\ref{fig-contacts-1D-crowding}B. For weaker
external confinement (larger $R$) we observe a mildly increasing dependence while
it is decreasing for the smaller cylinder radius. This behaviour indicates a
tradeoff between crowding and external confinement.

The effective overlap length of the rings along the cylinder axis is, in contrast,
a very reproducible function with the functional relation
\begin{equation}
\left< l_{AB}(\phi)\right>\approx\left< l_{AB}(0)\right>(1-\phi).
\label{eq-ideal-mix}
\end{equation}
of the crowding fraction $\phi$, compare Fig.~\ref{fig-contacts-1D-crowding}A.
This fact indicates a nearly ideal mixing of polymer monomers and crowding
particles, as if the chain connectivity plays a minor role. The absolute values
of $l_{AB}$ for different cylinder radii vary only marginally. The decrease in
Eq.~(\ref{eq-ideal-mix}) can be understood from a shrinkage of individual ring
polymers by the crowders, as rationalised for longitudinal ring dimensions in
Fig.~\ref{fig-RG2-single}B. We note a relatively small value of the overlap length
at all crowding densities used in simulations. It is consistent with the results
of Ref.~\cite{stasiak13} where, in the absence of crowders, a very limited
inter-penetration and overlap of the two polymer rings was obtained.

\begin{figure}
\begin{center}
\includegraphics[width=7cm]{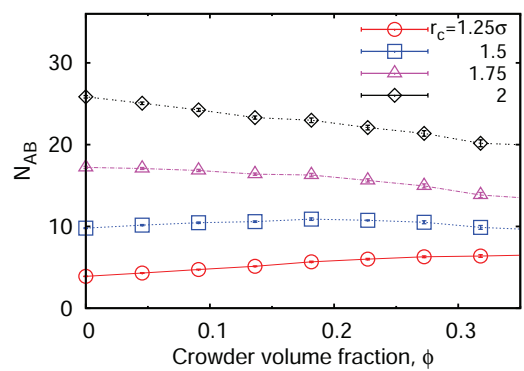}
\caption{Average number of ring-ring contacts computed for varying contact
distances $r_c$ at cylinder radius $R=4.5\sigma$. Other parameters are the
same as in Fig.~\ref{fig-NAB-crazy}.}
\label{fig-rc-all}
\end{center}
\end{figure}

We also systematically examined the effect of the contact distance $r_c$, defining
the overlap of both polymer monomers and crowders, on the number of ring-ring
contacts established in the simulations. We find that the average number of
contacts naturally grows with $r_c$, compare Fig.~\ref{fig-rc-all}. We also note
that the error bars somewhat increase with $\phi$ and $r_c$ but always stay smaller
than the symbol size. Here and below, as proposed in Ref.~\cite{stasiak13}, for the
ring-ring contacts the error bars are computed with the blocking method
introduced for correlated data sets in Ref.~\cite{blocking}.

\begin{figure}
\begin{center}
\includegraphics[width=16cm]{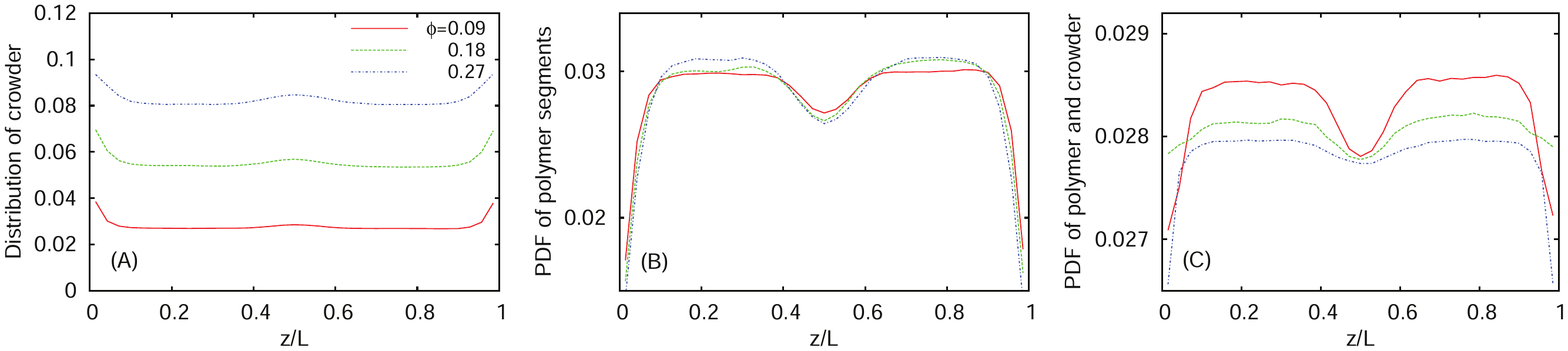}
\caption{Distribution of crowding particles (A), polymer monomers (B), and the
cumulative crowder-polymer distribution (C), plotted along the cylinder axis for
varying $\phi$ fractions. Parameters are the same as in Fig.~\ref{fig-NAB-crazy}.}
\label{fig-pdf}
\end{center}
\end{figure}

The statistical effects of the polymer-crowder mixing are analysed in terms of
their distributions in the simulation cell. We find that for weak and moderate
crowding fractions there is an accumulation of crowders near the cylinder ends,
as evidenced in Fig.~\ref{fig-pdf}A. In turn, the polymer monomers are located
preferentially off the middle of the cylinder, see Fig.~\ref{fig-pdf}B. We also
observe that at small $\phi$ the crowding particles are effectively excluded
from regions occupied by the polymers, thus facilitating ring-ring contacts. At
stronger crowding, a peak of crowding particles in the middle of the cell (between
the polymer rings) emerges, see the blue curve in Fig.~\ref{fig-pdf}A. These
mid-positioned crowding particles trigger an entropy-driven segregation of
polymer rings, and their three-dimensional contact number $N_{AB}$ decreases at
larger $\phi$ values (Fig.~\ref{fig-contacts-final}B). 

The mixing properties of polymers and crowders can be probed by the cumulative
probability distribution of their monomers shown Fig.~\ref{fig-pdf}C. The
decrease of $N_{AB}(\phi)$ at high $\phi$ is both due to a progressive emergence
of crowders in between the polymers and a longitudinal shrinkage of each of the
rings with $\phi$. The ideal $(1-\phi)$ polymer-crowder mixing is realised at
high $\phi$, when the sum of the distributions of the polymer monomers and crowders
is almost constant throughout the simulation cell, see Fig.~\ref{fig-pdf}C. Note
that in order to sample the polymer configurations equally well at varying
crowding fractions, longer simulation times are usually required at high $\phi$
values.

\subsection{Variation of the polymer length}

\begin{figure}
\begin{center}
\includegraphics[width=14cm]{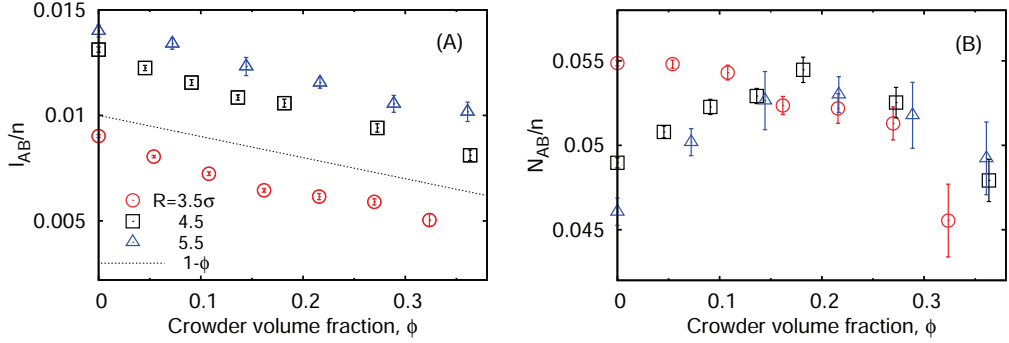}
\caption{Relative overlap length (A) and contact number of polymer rings (B) in
expanding simulation cells. In this figure the polymer volume fraction is kept
constant at $\phi_p\approx 0.09$, and the length of the polymer rings is $n=102$,
$200$, and $348$ monomers for cylinder radii of $R=3.5\sigma$, $4.5\sigma$,
$5.5\sigma$, respectively. Note the larger deviations around the mean for the
contact numbers $N_{AB}$ at more crowded conditions (the error bars are computed
by the blocking method \cite{blocking}). The square symbols represent the same
data set as in Fig.~\ref{fig-contacts-1D-crowding}B for $R=4.5\sigma$.}
\label{fig-contacts-final}
\end{center}
\end{figure}

In the previous sections, most of the results presented were obtained for a
constant polymer length of $n=200$ monomers. To be able to scale up the our
computations, we computed the ring-ring overlap length and contact number for
longer polymers and larger simulation cells. The geometrical proportions of the
confining cylinder, the aspect ratio $R/L$, were kept constant and at the same
time the ring length was adjusted so the polymer volume fraction $\phi_p=V_p/V$
stays constant at $\phi_p\approx0.09$. We observe that both the one-dimensional
overlap length of the polymer rings and their three-dimensional contact number
follow universal curves, after normalisation with respect to the number of
monomers, i.e., for $l_{AB}/n$ and $N_{AB}/n$. These results are illustrated in
Fig.~\ref{fig-contacts-final}, which is the main result of this work. The
ring-ring overlap length follows the $(1-\phi)$-asymptote typical for the ideal
mixing for all the parameters analysed in our simulations. In contrast, the
number of contacts $N_{AB}$ reveals a more delicate dependence. For small confining
cylinders the contact number is a monotonically decreasing function (red symbols
in Fig.~\ref{fig-contacts-final}B). For larger simulation cells, the value of
$N_{AB}(\phi)$ exhibits a maximum at $\phi\approx0.2$ (black and blue symbols in
Fig.~\ref{fig-contacts-final}B). This fraction $\phi\approx20\%$ is reminiscent of
the turning-point value for the non-monotonic dependencies on the fraction of
crowding molecules mentioned in the Introduction, namely, of the dimensions of
self-avoiding polymers \cite{schzleifer10} and diffusion-limited chemical
reactions \cite{kondev09}.

To compute $l_{AB}/n$ and $N_{AB}/n$, we averaged over $M=2\times10^5$, $2\times
10^5$, and $8\times10^4$ simulation steps for cylinder radii $R=3.5\sigma$,
$4.5\sigma$, $5.5\sigma$, respectively. The simulation time on a standard 3-3.5
GHz core machine for each crowding fraction $\phi$ presented in
Fig.~\ref{fig-contacts-final} is about 2, 3, and 10 days, respectively. To
accumulate reliable statistical information about the ring-ring contacts at
relatively large crowding fractions, particularly long simulations are required
because of the slower dynamics of inter-ring mixing. Last, at the same volume
fraction of crowders $\phi$, one can expect crowding particles of larger sizes
to cause stronger effects on mixing properties of polymer rings.

\section{Conclusions}
\label{sec-conclusions}

Based on extensive Langevin dynamics simulations we analysed the behaviour of
polymer chains of a circular topology in the presence of external confinement in
a cylindrical geometry and internal crowding by molecular crowding agents. The
size of the cylindrical confinement with respect to the monomer size and
the length of the chains was chosen to represent the situation of two DNA rings
in a typical bacillus cell. The crowding agents were represented by thermally
agitated, off-lattice mono-disperse hard spheres. We found that the topological
constraints restricting the polymer dynamics alter the response of partially
intermingled circular polymers compared to linear chains. In highly
crowded environments the polymer dynamics was demonstrated to be slowed down
significantly. We found that high concentrations of crowding agents facilitate the
spatial separation of ring polymers in cylindrical confinement. In addition,
we quantified the extent to which the presence of crowding agents slows down the
dynamics of the polymer-crowder system. The simulations for chains of varying
length demonstrates that our model results are robust and in principle scalable to
the dimensions of real bacterial cells. 

The effect of molecular crowding obtained above are applicable to de-mixing of
genome-sized DNA molecules inside bacterial cells as well as to the behaviour of
relatively short DNA plasmids confined in natural compartments inside eukaryotic
cells. The abundance of macromolecular crowders also offers a robust and
non-specific way to tune the amount of DNA-DNA contacts. The dynamics and
spatial occurrence of the latter are
vital for biological processes such as DNA-DNA recognition and DNA homologous
recombination \cite{stasiak13}, when the search for the homologous DNA partner
in a coil of a long DNA is to be performed. Note that it would be interesting
to analyse how the DNA-polymer segregation takes place in bacteria with other
than rod-like shapes, such as in nearly planar squarish or spherical bacteria,
see the discussion in Ref.~\cite{jun10}. 

The compartmentalisation of obstacles and polymer chains we observed can also
have implications for ageing phenomena of bacterial cells. For \emph{E. Coli\/}
cells, for instance, a localisation of age-related protein aggregates in
low-crowding regions near the cell poles and in between of the two DNA nucleoids
was observed. This effect was recently quantified by computer simulations at
higher degrees of polymer and DNA crowding inside the nucleoids that hinder
the diffusion of these protein aggregates \cite{hugues}.

\ack

JS thanks W. K. Kim for providing a computer code for analysis of ring swapping
dynamics.
The authors acknowledge funding from the Academy of Finland (FiDiPro scheme
to RM), the German Research Council (DFG Grant CH 707/5-1 to AGC), and the
German Federal Ministry for Education and Research (to JS).


\section*{References}

\begin{thebibliography}{99}

\bibitem{wright10entropy} S. Jun and A. Wright, Nature Rev. Microbiol. \textbf{8}, 600 (2010).

\bibitem{hancock-eu-crowding} R. Hancock, PLoS ONE \textbf{7}, e36045 (2012).

\bibitem{mirny1} J. Dekker, A. Marti-Renom, and L. A. Mirny, Nature Rev. Genet.
\textbf{14}, 390 (2013).

\bibitem{mirny2} T. B. K. Lee et al., Science \textbf{342}, 731 (2013).

\bibitem{kleckner-ecoli} D. Bates and N. Kleckner, Cell \textbf{121}, 899 (2005).
 
\bibitem{cremercremer} T. Cremer and C. Cremer, Nature Rev. Genet. \textbf{2}, 292 (2001).

\bibitem{stasiak09} J. Dorier and A. Stasiak, Nucl. Acids Res. \textbf{37}, 6316 (2009).

\bibitem{rosaeveraers08} A. Rosa and R. Everaers, PLoS Comput. Biol. \textbf{4}, e1000153 (2008).

\bibitem{rosaeveraers11} A. Rosa, N. B. Becker and R. Everaers, Biophys. J. \textbf{98}, 2410 (2010).

\bibitem{yuval} I. Bronstein, Y. Israel, E. Kepten, S. Mai, Y. Shav-Tal,
E. Barkai, and Y. Garini, Phys. Rev. Lett. \textbf{103}, 018102 (2009).

\bibitem{kroy} K. Kroy and J. Glaeser, New J. Phys. \textbf{9}, 416 (2007).

\bibitem{phage09} D. Marenduzzo et al., Proc. Natl. Acad. Sci. U.S.A. \textbf{106}, 22269 (2009).

\bibitem{virus13}  G. Saper et al., Nucl. Acids Res. \textbf{41}, 1569 (2013).

\bibitem{maarel} C. Zhang et al.,  Proc. Natl. Acad. Sci. U.S.A. \textbf{106}, 16651 (2009).

\bibitem{zimm96} S. B. Zimmerman and L. D. Murphy, FEBS Lett. \textbf{390}, 245 (1996).

\bibitem{elk10} A. R. McGuffee and A. H. Elcock, PLoS Comp. Biol. \textbf{6}, e1000694 (2010).

\bibitem{minton93} S. B. Zimmerman and A. P. Minton, Annu. Rev. Biophys. Biomol. Struct.\textbf{ 22}, 27 (1993).

\bibitem{igor} I. Goychuk, Adv.  Chem. Phys. \textbf{150}, 187 (2012).

\bibitem{jeon1} J.-H. Jeon, H. Martinez-Seara Monne, M. Javanainen, and R.
Metzler, Phys. Rev. Lett. \textbf{109}, 188103 (2012).\\
J-.H. Jeon, N. Leijnse, L. Oddershede, and R. Metzler, New J. Phys. \textbf{15},
045011 (2013).

\bibitem{pt} E. Barkai, Y. Garini, and R. Metzler, Physics Today \textbf{65}(8),
29 (2012).\\
F. H{\"o}fling and T. Franosch, Rep. Prog. Phys. \textbf{76}, 046602 (2013).

\bibitem{golding} I. Golding and E. C. Cox, Phys. Rev. Lett. \textbf{96} 098102
(2006).

\bibitem{weber} S. C. Weber, A. J. Spakowitz, and J. A. Theriot, Phys. Rev.
Lett. \textbf{104}, 238102 (2010).

\bibitem{weiss} J. Szymanski and M. Weiss, Phys. Rev. Lett. \textbf{103},
038102 (2009).

\bibitem{jeon} J.-H. Jeon, V. Tejedor, S. Burov, E. Barkai, C. Selhuber-Unkel, K.
Berg-S{\o}rensen, L. Oddershede, and R. Metzler, Phys. Rev. Lett. \textbf{106},
048103 (2011).

\bibitem{tabei} S. M. A. Tabei, S. Burov, H. Y. Kim, A. Kuznetsov, T. Huynh,
J. Jureller, L. H. Philipson, A. R. Dinner, and N. F. Scherer, Proc. Natl.
Acad. Sci. USA \textbf{110}, 4911 (2013).

\bibitem{zhou04}  H. X. Zhou, J. Mol. Recognit. \textbf{17}, 368 (2004).

\bibitem{kapral12} C. Echeverria and R. Kapral, Phys. Chem. Chem. Phys.  \textbf{14}, 6755 (2012).

\bibitem{marenduzzo13} C. A. Brackley, M. E. Cates, and D. Marenduzzo, Phys. Rev.
Lett. \textbf{111}, 108101 (2013).

\bibitem{weiss13crowd} O. Stiehl, K. Weidner-Hertrampf, and M. Weiss, New J.
Phys. \textbf{15},  113010 (2013).

\bibitem{lang11} T. Kuhn et al., PLoS One \textbf{6}, e22962 (2011).

\bibitem{jun10} S. Jun, "Polymer Physics for Understanding Bacterial Chromosomes", Chapter 6 in "Bacterial Chromatin", p. 97-116, Eds.: R. T. Dame and C. J. Dorman, Springer, (2010).

\bibitem{mulder-repulsion} S. Jun and B. Mulder, Proc. Natl. Acad. Sci. USA
\textbf{103}, 12388 (2006), and references cited therein.

\bibitem{metzler02links} R. Metzler, Y. Kantor, and M. Kardar, Phys. Rev. E
\textbf{66}, 022102 (2002).

\bibitem{rubin-rings-obstacles} M. Rubinstein, Phys. Rev. Lett. \textbf{57}, 3023
(1986).

\bibitem{ital-plane} C. Micheletti and E. Orlandini, Macromol. \textbf{45}, 2113
(2012).


\bibitem{ital-sphere} L. Tubiana et al., Phys. Rev. Lett. \textbf{107}, 188302
(2011).

\bibitem{ital-review} C. Micheletti, D. Marenduzzo, and E. Orlandini, Phys. Rep. \textbf{504}, 1 (2011).

\bibitem{koreans12} Y. Jung et al., Soft Matter \textbf{8}, 2095 (2012).

\bibitem{stasiak14chrom-NAR} F. Benedetti, J. Dorier, Y. Burnier, and A.
Stasiak, Nucl. Acids Res., at press; DOI:10.1093/nar/gkt1353.

\bibitem{likos12} A. Narros et al., Macromol. \textbf{46}, 3654 (2013).

\bibitem{reich13rings} S. Y. Reigh and D. Y. Yoon, ACS Macro Lett. \textbf{2}, 296 (2013).

\bibitem{slovak13} D. Racko and P. Cifra, J. Chem. Phys. \textbf{138}, 184904 (2013).

\bibitem{stasiak13} J. Dorier and A. Stasiak, Nucl. Acids Res. \textbf{41}, 6808 (2013).

\bibitem{italy-gel-rings} D. Micheletti et al., arXiv:1306.4965

\bibitem{orlandini10pre} E. Orlandini et al., Phys. Rev. E \textbf{82}, 050804(R) (2010).

\bibitem{note-ring-RG-scaling} Two scaling regimes for ring dimensions were
predicted in Ref.~\cite{orlandini10pre}. Namely, rings smaller than the lattice
size $b$ behave as a self-avoiding walk, $\left< R_g^2 \right>\sim n^{1.15}$, while
large rings follow the law for the branched polymers [G. Parisi and N. Sourlas,
Phys. Rev. Lett. \textbf{46}, 871 (1981)]: $\left< R_g^2 \right>\sim n^1$.
Similarly, for knotted rings the dynamics turns from a self-avoiding walk to a
branched polymer dynamics \cite{orlandini10pre}.

\bibitem{schzleifer11} J. S. Kim et al., Phys. Rev. Lett. \textbf{106}, 168102 (2011).

\bibitem{schzleifer10} J. S. Kim and I. Szleifer, J. Phys. Chem. C \textbf{114}, 20864 (2010).

\bibitem{kondev09} J. D. Schmit, E. Kamber, and J. Kondev, Phys. Rev. Lett. \textbf{102}, 218302 (2009).

\bibitem{levi-crowd} A. Marcovitz and Y. Levy, Biophys. J., \textbf{104} 2042 (2013). 

\bibitem{kim07transl} A. Gopinathan and Y. W. Kim, Phys. Rev. Lett. \textbf{99}, 228106 (2007).

\bibitem{liu-crowd} Y. Chen and K. Luo, J. Chem. Phys. \textbf{138}, 204903 (2013).

\bibitem{tere13} A. Lappala, A. Zaccone, and E. M. Terentjev, Scientific Reports \textbf{3}, 3103 (2013).

\bibitem{shin-looping} J. Shin, A. G. Cherstvy, and R. Metzler, work in preparation.

\bibitem{scr13virus} J. Shin, A. G. Cherstvy, and R. Metzler,  arXiv:1310.5531.

\bibitem{voloDNAdiameter2} V. V. Rybenkov, A. V. Vologodskii, and N. R. Cozzarelli, Nucl. Acids Res. \textbf{25}, 1412 (1997).

\bibitem{elcock-ecoli-cytoplasm} A. R. McGuffee and A. H. Elcock, PLoS Comp. Biol. \textbf{6}, e1000694 (2010).

\bibitem{upshift} J. T. Mika et al, Mol. Microbiol. \textbf{77}, 200 (2010).

\bibitem{dna-elst} A. Savelyev, C. K. Materese, and G. A. Papoian, J. Am. Chem. Soc. \textbf{133}, 19290 (2011).

\bibitem{ac11rings} A. G. Cherstvy, J. Biol. Phys. \textbf{37}, 227 (2011).

\bibitem{ac11rings2} A. G. Cherstvy, J. Phys. Chem. B \textbf{115}, 4286 (2011).

\bibitem{skolnick-hydro} T. Ando and J. Skolnick, Proc. Natl. Acad. Sci. U.S.A. \textbf{107}, 18457 (2010).

\bibitem{wensink-hydro} H. H. Wensink et al.,  Proc. Natl. Acad. Sci. U.S.A. \textbf{109} 14308 (2012).

\bibitem{china-rings-hydro} G. A. Hegde et al., J. Chem. Phys. \textbf{135}, 184901 (2011).

\bibitem{korea09scaling} Y. Jung et al., Phys. Rev. E \textbf{79}, 061912 (2009).

\bibitem{prot-dna} A. G. Cherstvy and V. B. Teif, J. Biol. Phys. \textbf{39}, 363 (2013).

\bibitem{winkler-shear} R. G. Winkler, Phys. Rev. Lett. \textbf{97}, 128301 (2006).

\bibitem{winkler-shear2} C. C. Huang et al., J. Phys.: Cond. Matt. \textbf{24}, 284131 (2012).

\bibitem{chuang} J. Chuang, Y. Kantor, and M. Kardar, Phys. Rev. E \textbf{65},
011802 (2001).\\
K. Luo, T. Ala-Nissila, S.-C. Ying, and R. Metzler, Europhys. Lett. \textbf{88},
68006 (2009).

\bibitem{blocking} H. Flyvbjerg and H. G. Petersen, J. Chem. Phys. \textbf{91}, 461 (1989).

\bibitem{hugues} A. S. Coquel et al, PLoS Comput. Biol. \textbf{9}, e1003038 (2013).

\end{thebibliography}
\end{document}